\RequirePackage{fix-cm}
\documentclass[smallextended]{svjour3}       
\smartqed  
\usepackage{graphicx}
\usepackage{amssymb}
\journalname{Journal of Low Temperature Physics}
\begin{document}

\title{Recent progress in quantum simulation using superconducting circuits
\thanks{Financial support from the Academy of Finland (project 263457, and the Center of Excellence ``Low Temperature Quantum Phenomena and Devices'' project 250280) and FQXi is gratefully
acknowledged.}
}


\author{G. S. Paraoanu}


\institute{G. S. Paraoanu \at
              O. V. Lounasmaa Laboratory, \\
              Aalto University School of Science \\
               P.O. Box 15100, FI-00076, Finland
}

\date{Received: date / Accepted: date}

\maketitle

\begin{abstract}
Quantum systems are notoriously difficult to simulate with classical means.
Recently, the idea of using another quantum system - which is experimentally more controllable - as
a simulator for the original problem has gained significant momentum. Amongst the experimental platforms studied as quantum simulators, superconducting qubits are one of the most promising, due to relative straightforward scalability, easy design, and integration with standard electronics. Here I review the recent state-of-the art in the field and the prospects for simulating systems ranging from relativistic quantum fields to quantum many-body systems.
\end{abstract}

\section{Introduction}
\label{intro}

The idea of extracting information about one quantum system by mathematically mapping it to another one which is more easily accessible goes back to Richard Feynman \cite{Feynman} and David Deutsch \cite{deutsch}. The development of the field of quantum simulation is intertwined with that of quantum computing, since from the beginning quantum simulation was envisioned as the main application of a quantum computer. A universal quantum computer would be able to efficiently simulate the dynamics of quantum systems, provided that the Hamiltonian is built up from local interactions \cite{lloyd}. Powerful mathematical results such as the threshold theorem and quantum error-correcting codes suggest that quantum processors can be made robust against noise and imperfections - therefore fault-tolerant quantum computing is in principle possible \cite{nielsen}. The threshold theorem shows that running a quantum algorithm with a finite probability of error can be done on a hardware with noisy or faulty components, provided that the probability of failing for each component remains below a threshold value. This demonstrates that it is possible to prevent the accumulation and propagation of errors when running a quantum algorithm. However, the requirements set by the threshold theorem and the overhead in the number of qubits for error corrections for realizing a quantum processor are a tall order for the applied physicist. In spite of significant advances on the experimental front, it has been gradually realized that progress in the direction of gate-based universal quantum computing is probably going to be a long-term effort \cite{physperspect}. As a response to this challenges, the field of quantum simulations proposes a more pragmatic strategy: one searches for problems in quantum physics that can be mapped onto another mathematical model, which in turn can be straightforwardly implemented with the existing hardware. It is hoped that the requirements for simulating these models efficiently and accurately might not be so drastic. As shown in this review, quantum physics offers a plethora of difficult and important problems that are amenable to this approach. The appeal of the field of quantum simulations is twofold: first, it is hoped that quantum simulators might provide a route towards efficient solving of problems and models which are difficult to implement on classical computers; second, they illuminate powerful and scientifically fertile analogies and mathematical mappings between seemingly unrelated phenomena, ranging from solid-state physics and chemistry to particle physics and cosmology.

A variety of systems (trapped ions, atoms in optical lattices, polar molecules, quantum dots, nuclear spins, superconducting circuits {\it etc.}) have been recognized as promising candidates for quantum simulation \cite{buluta1,buluta2,ciraczoller,georgescu,lewenstein,lewensteinM}. Not surprisingly, these systems are the same as those envisioned for the implementation of quantum computers. The field of simulations thus has a start-up advantage, being able to heavily capitalize on the technological advances in coherence time, addressability,  tunability {\it etc.} achieved by the various implementations of few-qubits quantum processors.

In the case of superconducting circuits and qubits, there have been remarkable technical achievements in design, fabrication, operation, and measuring techniques \cite{shnirman,devoret,clarke,wendin,girvinrev,nori,schoelkopf}. Superconducting qubits are mesoscopic element circuits that behave like artificial atoms, having quantized energy levels which can be excited by microwave fields. While in the case of atom/ion-based qubits the energy levels are due to the motion of electrons around the nucleus, in the case of superconducting qubits the energy levels are created by the collective motion of electrons in the superconductor and through Josephson links, either as electron plasma oscillations (charge qubit, transmon, quantronium) or as persistent currents (phase qubit, flux qubit). Differently from atom/ion- based qubits, the superconducting qubits interact strongly with electromagnetic fields, which is both a blessing ({\it e.g.} reaching the strong coupling limit of cavity QED is easy) and a curse - since the undesired interaction with the electromagnetic modes of the environment result in shorter decoherence times. There exists already a quite large toolbox of techniques available for the experimentalist working with these circuits. The qubits can be coupled capacitively or inductively, by the design of the circuit. The interaction between qubits can be tuned by applying pulses that put the qubits on- and off- resonance \cite{cnotmartinis,singleshotesteve}, by using a third off-resonant qubit or resonator \cite{niskanen,majer}, by employing specially-designed current-controlled coupling circuits \cite{coherentcoupling}, or by using the sidebands resulting from applying microwaves \cite{mw1,mw2,mw3,mw4,mw5}. A lot of progress has been done in the use of novel measurement techniques, either specific to the physics of Josephson junctions ({\it e.g.} switching techniques \cite{sw1,sw2,sw3,sw4,sw5,sw6}) or inspired from quantum optics ({\it e.g.} homodyne measurements \cite{transmon1,transmon2}). Various extensions of these methods include partial and interaction-free measurements \cite{partial1,partial2,partial3,partial4,partial5,partial6}, single-shot measurements using the cavity at large photon numbers \cite{singleshot} and bifurcation amplifiers \cite{bif}, and measurement feedback enabled by the use of parametric amplifiers \cite{feedback1,feedback2,feedback3,feedback4}.

 A convenient classification of simulators is into analog and digital, depending on weather the time evolution of the simulator is continuous in time or, respectively, in discrete steps (stroboscopic). The digital version uses the standard model of quantum circuits based on quantum gates. The simulation of dynamical evolution discussed below is an example of digital simulation. The technical difficulty of this approach is thus the same as generally with quantum computing (the need of large number of well-controlled gates) and the bounds on decoherence, fidelities of state preparation, gate errors, and measurement errors are set by the overall desired precision of the calculation; they can be improved by the usual techniques of quantum error correction and fault tolerance.  The analog version is the emulation of the system under investigation by another quantum system, realizing the same physics or a mathematically equivalent form of it using continuously-tunable parameters in a better controlled setup. The analog quantum simulators realized with superconducting circuit elements can be alternatively seen as realizations of quantum metamaterials with externally-adjustable material characteristics \cite{zagoskinmetamaterials1,zagoskinmetamaterials2}. So far, there are two main categories of devices that can be subscribed to the class of analog simulators: the first are adiabatic and quantum annealing simulators, while the second class is that of quantum hardware emulators. The first class realizes a form of quantum computation based on the adiabatic theorem - a system with parameters slowly changing in time will stay in the ground state. The idea is to access nontrivial many-body ground states by starting in a known simple ground state and varying the parameters of the quantum processor Hamiltonian \cite{adiabatic}. The second class of analog simulators are essentially dedicated devices designed to solve a challenging problem by recreating the same physics in a simpler, easily-accessible, or more precisely controlled experiment. For hardware emulators there is no threshold theorem and no general error correction techniques are known: the effect of decoherence, measurement fidelities, {\it etc.} has to be studied for each simulator separately. How to characterize the efficiency and reliability against imperfections of analog simulators is an open theoretical issue. The niche for analog quantum simulators might be an intermediate regime of noise - where the noise level is too large for fault-tolerant universal quantum computing, but still small enough to outperform classical computers \cite{lewenstein}. With the levels of decoherence being constantly reduced over the last decade, superconducting devices and circuits are excellent candidates for analog quantum simulators.

 In this review I will not follow the digital/analog classification - instead, the simulators are categorized
 according to the problem they can solve or to the system they are meant to emulate. From simulations of gauge fields in lattices to condensed matter systems such as frustrated spins and spin liquids, one may expect that novel phases of matter can be discovered and new insights can be gained about the physics of these systems.

\section{Digital simulators of dynamical evolution}

 The first demonstration that the quantum simulation of the dynamics of a manybody system can be more efficient than the classical computer simulation is due to Lloyd \cite{lloyd}. He showed that the dynamics of a quantum many-body system can be simulated by slicing the time evolution into a sequence of operations realized with local Hamiltonians. Suppose the total Hamiltonian can be written as
\begin{equation}
H=\sum_{k=1}^{N}H_{k},
\end{equation}
where $H_{k}$ are local Hamiltonians. For example, for the quantum Heisenberg model
$H = -\frac{1}{2} \sum_{j=1}^{N} \left(J_{x}\sigma^{x}_{j}\sigma^{x}_{j+1}+J_{y}\sigma^{y}_{j}\sigma^{y}_{j+1}+
J_{z}\sigma^{z}_{j}\sigma^{z}_{j+1} - h\sigma_{j}^{z}\right)$, where $J_{x,y,z}$ are interaction (coupling) constants and $h $ is the external magnetic field. In this case, the $H_{k}$'s are the interactions between two nearby lattice spins $j$ and $j+1$, $\sigma^{x}_{j}\sigma^{x}_{j+1}$, $\sigma^{y}_{j}\sigma^{y}_{j+1}$, as well as the single-spin Pauli operators $\sigma^{z}_{j}$.
Then, the evolution of the system under $H$ can be approximated as
\begin{equation}
e^{\frac{i}{\hbar}Ht} \approx \left(e^{\frac{i}{\hbar}H_1 t/n}e^{\frac{i}{\hbar}H_2 t/n} ...  e^{\frac{i}{\hbar}H_N t/n}\right)^{n}.\label{expa}
\end{equation}
where the error can be made as small as needed by taking a large enough number of steps $n$. This result follows in a similar way as the two-operator Trotter formula, ${\rm lim}_{n\rightarrow \infty}\left(e^{iAt/n}e^{iBt/n}\right)^{n} = e^{i(A+B)t}$ \cite{trotter1,trotter2}. It is important to note that this expansion holds even if $H_{1}, H_{2}, ..., H_n$
do not commute.

The expansion Eq. (\ref{expa}) makes the simulation efficient because now each of the operators $H_{k}$ acts on a reduced Hilbert space.
For example, for the quantum 1D Heisenberg model defined on $N$ sites as above, the single-qubit terms act on a 2-dimensional Hilbert space, while the interaction terms act on a 4-dimensional Hilbert space. The simulation of the evolution operators $\exp (\frac{i}{\hbar}H_{k} t/n)$ is given by the number of entries when these operators are written in matrix form, and it is 4 for single-qubit terms and 16 for two-qubit terms. The total number of operations in the simulation is therefore $\approx nN(4+3\times 16)=52nN$, with $n$ here controlling the error - the larger $n$ the smaller the error. Contrast this with the case of a classical simulation of the same problem where, in order to directly calculate the evolution $\exp (-\frac{i}{\hbar}Ht)$, one needs to exponentiate a $2^{N}\times 2^{N}$ matrix.

It is in principle possible to realize this type of digital quantum simulation by using superconducting  circuits. All the needed components exist, and implementations of quantum gates have been already demonstrated with various types of superconducting qubits \cite{gates1,gates2,gates3,gates4}. Specific constructions for the gate sequence implementing the Trotter decomposition have been constructed for a chain of transmons (all coupled to a resonator) that can simulate the Heisenberg and frustrated Ising models with two and three spins \cite{wallraff}. In general, a high accuracy would require the use of fault-tolerant error correction; this raises the delicate question of the amount of resources needed, which unfortunately tend to grow exponentially with the degree of precision we wish to impose \cite{chuang}. However, if the requirements on precision are not too strong, the digital simulation might still perform better than the classical computation, especially for large number of spins.

\section{Single-qubit simulators}

One usually thinks of a simulator as a system consisting of many quantum-level components (artificial atoms), but it is possible to take a minimalist approach and attempt to use a  single artificial atom as a simulator. Several interesting experiments have already been done by following this strategy.

\subsection{Simulating a large spin with a multilevel artificial atom}

Multilevel systems are interesting due to effects such as electromagnetically induced transparency, coherent population trapping, and the Autler-Townes effect.  These effects have been already demonstrated experimentally with superconducting qubits \cite{AT1,AT2,AT3,AT4}. Multilevel systems can also be operated as quantum simulators of large spins. To emulate a spin $s$, one needs a 2s+1 dimensional Hilbert space on which a basis is defined as $\{|s,m\rangle\}$ with $m=-s, -s+1,..., s$. Thus a quantum system with $d=2s+1$ degrees of freedom is suitable. This allows the analog simulation of the dynamics of spins with different quantum numbers (spin-1/2, 1, and 3/2) by employing the first five levels of a phase qubit \cite{Neely}. For example, a spin-1/2 is emulated by using two levels, a spin-1 by using 3 levels, and a spin-3/2 by using 4 levels. In this way one can study the quantum dynamics of large spins as well as the associated geometric phases.

More precisely, the experiment \cite{Neely} relies on the analogy between the Hamiltonian of a multilevel system coupled to an external field and the generators of rotation around the $x$-direction for the $s$-spin. Any superconducting qubit is a multilevel system; when this system interacts with microwaves, the Hamiltonian is of the generic form
\begin{equation}
H = \sum_{n} E_{n}|n><n| + \hbar\sum_{n}g_{n}(t) \left[|n+1><n| + |n><n+1|\right],
\end{equation}
where $g_{n}(t) = \sum_{m}g_{n,m} \cos (\omega_{m}t)$ contains all the $m$ components of  the applied microwave tone, which couple to the qubit's transitions $n\rightarrow n+1$ with strengths $g_{n,m}$.
The Schr\"odinger equation $i\hbar (d/dt) |\psi (t)\rangle = H |\psi (t)\rangle$ can be most easily solved in a multiple-rotating frame defined by a unitary $U(t)$, where it reads  $i\hbar (d/dt) |\tilde{\psi} (t)\rangle = \tilde{H} |\tilde{\psi} (t)\rangle$, with $|\tilde{\psi} (t)\rangle = U^{\dagger}(t)|\psi\rangle$ and $\tilde{H}(t) = U^{\dagger}(t)H(t)U(t) + i \hbar (d U^{\dagger}(t)/dt) U(t)$. If we now take
\begin{equation}
U(t) = e^{-\frac{i}{\hbar}\sum_{n}E_{n}t |n><n|},
\end{equation}
and take each of the tones $m$ exactly resonant to a transition $n\rightarrow n+1$, by performing a rotating wave
approximation we get \cite{jian_prb}
\begin{equation}
\tilde{H}_{\rm eff} = \sum_{n=0}\frac{\hbar g_{n,n}}{2}[|n><n+1| + |n+1><n|],
\end{equation}
where $g_{n,n}$ are the Rabi frequencies between consecutive transitions. The phase factors of the microwave field can be included as well in this calculation. Here we have assumed that we can neglect oscillating terms of the type $g_{n,m}\exp [-i\omega_{m}t + i(E_{n+1}-E_{n})t/\hbar)] |n+1><n|$ if $m\neq n$. This approximation is not valid in general and these cross-coupling  terms can produce measurable effects \cite{scirep}.

On the  other hand, the generators of rotation around the $x$ and $y$ -direction for spin-1/2, spin-1, and spin-3/2 are as follows:
\begin{eqnarray}
 \sigma_{x}&=&
 \left(\begin{array}{cc} 0&1\\ 1&0
\end{array}\right); \sigma_{y}=
 \left(\begin{array}{cc} 0&-i\\ i&0
\end{array}\right),
\\
J_{x}^{(1)}&=& \frac{1}{\sqrt{2}}
 \left(\begin{array}{ccc} 0&1&0\\ 1&0&1 \\ 0&1&0
\end{array}\right); J_{y}^{(1)}= \frac{1}{\sqrt{2}}
 \left(\begin{array}{ccc} 0&-i&0\\ i&0&-i \\ 0&i&0
\end{array}\right),
\\
J_{x}^{(3/2)}&=&
 \frac{1}{2}\left(\begin{array}{cccc} 0&\sqrt{3}&0&0\\\sqrt{3}&0&2&0 \\ 0&2&0&\sqrt{3} \\0&0&\sqrt{3}&0
\end{array}\right);
J_{y}^{(3/2)}=
 \frac{1}{2}\left(\begin{array}{cccc} 0&\-i\sqrt{3}&0&0\\i\sqrt{3}&0&-2i&0 \\ 0&2i&0&-i\sqrt{3} \\0&0&i\sqrt{3}&0
\end{array}\right).
\\
\end{eqnarray}
Thus, rotations around the $x$ and $y$ - axes can be realized simply by externally-controlled microwave fields, resulting in a time-dependent $g_{n,n}$.

\subsection{Simulating effects in noninteracting many-body systems}

In his foundational paper for the field of quantum simulation \cite{lloyd}, Lloyd noticed that, while for quantum computing decoherence is a liability, in the case of quantum simulation it can be used as an asset. Indeed, since  the simulated system is itself embedded in some environment, the easiest way to mimick its dynamics is by using an open system with known or controlled decoherence. For example, for a system of nuclear spins characterized by frequency $\omega_{\rm sys}$, decay time $T_{\rm sys, 1}$ and decoherence time $T_{\rm sys, 2}$, it is enough to use as simulator a two-level system with corresponding parameters $\omega_{0}$, $T_1$, and $T_2$, such that $\omega_{\rm sys}/\omega_{0} = T_{1}/T_{\rm sys, 1} =T_{2}/T_{\rm sys, 2}$. In this way, the quantum state of the spin at the moment $t$ is reproduced in the simulator at a time $t\omega_{\rm sys}/\omega_{0}$.

This general idea has been realized in a recent experiment \cite{motion} simulating an effect known especially in nuclear magnetic resonance (NMR) as motional averaging and narrowing. Atoms in materials can experience a different chemical environment in different regions of space, and due to their thermal energy they can randomly move in and out of these regions. In NMR experiments, this effect is typically studied by raising the temperature of the sample, thus accelerating the random motion of the atoms. Such a noninteracting ensemble of atoms can be simulated with only one qubit simply by dividing the total measurement time into time bins and emulating one atom in each of the bins. The effect of the random chemical potential has been emulated by randomly modulating the transition frequency \cite{motion}. The Hamiltonian for this simulator is
\begin{equation}
H = \frac{\hbar}{2}\left[\omega_{0} + \xi (t)\sigma_{z}\right] + \hbar g \cos (\omega t) \sigma_{x}, \label{hmmm}
\end{equation}
where $\xi(t)$ is a random variable with units of frequency, taking the values $\pm \xi$ and satisfying the time-correlations
\begin{equation}
\langle \xi (t) \xi (t+\tau ) \rangle_{\xi} = \xi^2 e^{-2\chi \tau},
\end{equation}
where $\chi$ is the characteristic  inverse-time switching scale and the last term in Eq. (\ref{hmmm}) is the drive. Dissipation is introduced in the simulator as a Lindblad superoperator.
Thus, in this experiment ensemble averages are mapped into time averages over the measurement periods, the temperature corresponds to the inverse switching time $\chi$, the values of the chemical potentials in the two regions of space correspond to $\pm \xi$, and the Larmor frequency of the nuclear spins $\omega_{\rm sys}$ and the times $T_{\rm sys, 1}$, $T_{\rm sys, 2}$ are mapped respectively into the qubit as $\omega_{0}$ and relaxation and dephasing times $T_{1}$, $T_{2}$.

\section{Simulators of relativistic fields and effects}

To observe analog relativistic effects such as Klein tunneling and {\it Zitterbewegung} in 1+1 dimensions is relatively simple: one needs a two-level systems and a continuous degree of freedom. A simple qubit-resonator circuit such as the one used for transmons can be used \cite{zitt}. A more difficult problem is solving problems in interacting relativistic quantum fields, and various systems have been proposed as simulators. The implementation of non-Abelian gauge theories with Josephson arrays has been discussed in Ref. \cite{discrete}. A new quantum algorithm for digital simulation of relativistic scattering processes for massive field theories with $\Phi^4$ self-interaction has been recently proposed \cite{preskill}.

\subsection{The massless Klein-Gordon field with tunable speed of propagation or boundary conditions: the dynamical Casimir effect and the  Hawking effect}

If an array of SQUIDs is used as the signal line of a coplanar waveguide transmission line \cite{pasi}, the Lagrangian density takes the form of the Klein-Gordon massless field,
\begin{equation}
{\cal L}[\partial_{t}\varphi , \partial_{x}\varphi ] = \frac{\rm C}{2}(\partial_{t}\varphi )^2 -\frac{1}{2{\rm L}}(\partial_{x}\varphi )^2,
\end{equation}
where $\varphi = \varphi (x,t)$ is the flux variable along the array, and ${\rm C}$ and ${\rm L}$ are the capacitance with respect to the ground per unit length and, respectively, the inductance per unit length. This yields the massless Klein-Gordon equation in  1+1 dimensions,
\begin{equation}
\Box \varphi =0,
\end{equation}
where the d'Alembert operator in 1+1 dimensions is $\Box =c^{-2}\partial^2/\partial t^2 - \partial^2/\partial x^{2}$, with $c=1/\sqrt{\rm LC}$ the propagation speed. Now, by using an external magnetic field, it is possible to make the inductance per unit length ${\rm L}$ dependent on both space and time, ${\rm L} (x,t)$. This makes also the speed $c$ dependent on space and time $c=c(x,t)$.

There are two ways of using this externally-controlled time- and space- dependence of the speed $c$ to create interesting effects:

i) The first option is to keep the time-dependence in a well-defined region of space, that is, $c(x,t)$ is varied in a fixed space interval defined on the sample by the proximity of a bias coil. This gives rise to the dynamical Casimir effect. In the dynamical Casimir effect, the change of the speed $c$  can be realized over a length either much smaller or of the same order as the wavelength. If it is realized over a length much smaller than the wavelength of the field, the process can be assimilated to a modulation a boundary condition \cite{wilson}. If the change is realized over a length of the same order as the wavelength, the process can be regarded as a modulation of the index of refraction \cite{pasi}.

ii) The second option is to have the space region over which $c$ is varied moving along the chain. This can be used to produce an analog of the Hawking effect, described in more detail later in this subsection.

The dynamical Casimir effect is a process by which the vacuum fluctuations of a field are transformed into real particles (typically photons) by the action of an external modulation. This modulation changes a parameter entering the Euler-Lagrange equations (or in the  Hamiltonian) of the system. The reason why this change creates particles is very general, and can be understood as follows: any quantum system (in general any quantum field) has a structure of energy levels as given by solving  the field equations. The ground state is by definition the lowest-energy state, to which we associate a zero number of particles. If one now changes the structure of the energy levels by modifying a parameter in the Hamiltonian or Lagrangian, the new ground state will be in general different from the initial one. By expanding the new ground state in the old basis, one can see that the new ground state, as seen from the point of view of the initial system, has components with a non-zero number of particles. The main experimental challenge to realizing the dynamical Casimir effect is that, in order to get a significant number of particles created, the parametric modulation should be nonadiabatic when compared to the energy level separation of the unperturbed system. Fast changes in materials and  Josephson devices embedded in electromagnetic cavities are particularly suited for this task, and early proposals have exploited precisely these properties \cite{early1,early2,early3}.

These processes can be implemented in superconducting circuits by modulating the inductance per unit length and thus the speed of light in SQUID-based arrays, realizing the first option i) mentioned above. In the dynamical Casimir experiment that uses modulation of the index of refraction \cite{pasi}, photons were generated at two frequencies that sum up to the pump (modulation) frequency. To measure the outgoing radiation the SQUID array was coupled into a transmission line by using a low-dissipation capacitor with vacuum gap fabricated with FIB (focused ion beam) \cite{pasi}. This formed a cavity with a quality factor Q = 50-100 and a resonant frequency tunable by an external magnetic field. The device was pumped at 10.8 GHz and the photons created by the dynamical Casimir effect were created at a frequency around 5.4 GHz.  By examining the covariance matrix extracted from the experimental data and employing standard criteria of non-separability for continuous variables, it was possible to prove that the photons created by the dynamical Casimir effect are in a non-separable two-mode squeezed state, as expected from theoretical considerations. In general, these correlations can be harvested and propagated further to other parts of a superconducting quantum circuit \cite{enta1,enta2}.

The dynamical Casimir effect can be realized also with the single-qubit simulators discussed in the previous section. One notices that the minimal requirements of realizing an emulation of the dynamical Casimir effect is to have a harmonic oscillator with a fast-tunable potential. With superconducting circuits, this can be done for example by using a double-SQUID where the shape of the landscape of potential energy of the circuit can be changed very fast by using rapid single flux quanta techniques \cite{zeilinger}.

 The  next important effect, considered to be a cornerstone of our understanding of quantum physics in the presence of  gravitation, is the Hawking effect. The simulation of the Hawking effect has been proposed using a SQUID array where the speed of light is modulated by a pulse propagating at the speed $u$ in a nearby bias line \cite{nation}, corresponding to the second option ii) mentioned above. This means that the  speed of the microwave photons in the SQUID array has a dependence both on time $t$ and coordinate $x$ along the array, of the form $c(x-ut)$. In the co-moving coordinates associated to the frame moving with speed $u$, this results in the field equation
\begin{equation}
\frac{1}{\sqrt{-g_{\rm eff}}}\partial_{\mu}\left(g^{\mu\nu}_{\rm eff}\sqrt{-g_{\rm eff}}\partial_{\nu}\varphi\right) = 0, \label{instiga}
\end{equation}
where $g^{\mu\nu}_{\rm eff}$ is
an effective metric
\begin{equation}
g^{\mu\nu}_{\rm eff} = \frac{1}{c^2}\left(\begin{array}{cc} 1 &-u \\ -u & u^2 - c^2\end{array}\right),
\end{equation}
and $g_{\rm eff} = {\rm Det}[g^{\mu\nu}_{\rm eff}]$. One now recognizes that Eq. (\ref{instiga}) is the standard way of writing the Klein-Gordon equation in a curved spacetime, and that the metric $g^{\mu\nu}_{\rm eff}$
is
analogous to the one obtained for massive non-rotating bodies in Painlev\'e-Gullstrand coordinates. The event horizon $h$ is formed whenever $|c(x-ut)| = |u|$, and, in the laboratory frame, it moves together with the pulse. Radiation of photons by the Hawking effect is then expected at the horizon. Note that the Hawking effect is a purely kinematic effect, thus it does not depend on the specific physical mechanism that produces the event horizon.

The equivalent surface gravity for this experiment is
\begin{equation}
g_{h}=c_{h}\left\vert\frac{\partial c}{\partial x}\right\vert_{h},
\end{equation}
where $c$ is the speed of light and the index $h$ stands for horizon, defined as the point where $|u| =|c(x-ut)|$,
where $u$ is the propagation speed of the pulse. The surface gravity has dimensions of acceleration - which justifies its name as an analog surface gravity, and the relation between the surface gravity at the horizon and Hawking's
temperature is
\begin{equation}
k_{\rm B}T_{\rm Hawking} = \frac{\hbar}{2\pi c_{h}}g_{h}. \label{Hawk}
\end{equation}
The ratio between the surface gravity and the speed of light at the horizon is a frequency,
and this is the quantity that governs the Hawking process. In this proposed experiment it would be possible to create a Hawking
temperature above 100 mK, enough to be detected. Let us assume that
the speed of light is changed by a factor of  20\% over a distance of the order of mm: this yields $g_{h}\approx 10^{19}$m/s$^2$, which results in a detectable Hawking radiation corresponding to a detectable value of 100 mK. Note that this experiment would create an effective gravitational acceleration 18 orders of magnitude larger than the gravitational acceleration at the surface of the Earth! It corresponds to a ``miniature'' black hole with a Schwarzschild radius of about 1.5 mm and a mass of $10^{24}$kg. These incredibly large numbers simply mean that, being so much stronger, electromagnetic fields
are much more effective than gravitation in generating quantum effects based on vacuum fluctuations.

\subsection{Toward relativistic quantum information}

The idea of simulating cosmological phenomena in the laboratory, especially in systems such as atomic Bose-Einstein condensates and superfluid He has been around for some time \cite{viser,volovik}, and superconducting circuits are an important recent system suitable for such experiments \cite{nationrmp}.

To illustrate the potential of superconducting circuits for relativistic and cosmological effects, consider for example the Unruh effect. The prediction is that the ground state of a field seen from a frame moving in the Minkowski vacuum with acceleration $a_{\rm Unruh}$ with respect to a stationary (laboratory) frame is a thermal state with temperature
\begin{equation}
k_{\rm B}T_{\rm Unruh} = \frac{\hbar}{2\pi c} a_{\rm Unruh}, \label{Unr}
\end{equation}
in other words a detector that moves with acceleration in empty space will detect particles. Note that Unruh's and Hawking's expressions for the temperature Eqs. (\ref{Hawk},\ref{Unr}) are identical,
with the noninertial acceleration $a_{\rm Unruh}$ replaced by the gravitational acceleration $g_{h}$, as expected from the equivalence principle of general relativity. As in the  case of Hawking's effect, an immediate estimation shows that in order to detect a temperature of 100 mK one needs to move the detector with an acceleration $a_{\rm Unruh}\approx 10^{19}$ m/s${^2}$. Achieving such accelerations in the lab with real detectors is technically unrealistic.
But consider now what happens when we look at the displacement of the electromagnetic field modes that are achieved in the dynamical Casimir effect experiment. Let us say that the amplitude of the displacement achieved in a chain of SQUIDs is of the order of 1 mm, and the modulation frequency is of the order of 10 GHz. This results in a maximum velocity of  $2\pi \times 10^{7}$ m/s and a maximum acceleration of $\approx 4\times 10^{18}$ m/s. The maximum velocity thus can reach values comparable or even above that of the speed of microwaves in  coaxial cables and transmission lines. This opens the way to realizing causally-disconnected (space-like separated) experiments. The accelerations obtained are close to the acceleration required to get an Unruh temperature
well above that of a typical dilution refrigerator (20 mK).

These experiments demonstrate the potential of superconducting quantum circuits to serve as a platform for simulating effects from cosmology and quantum field theory. One can go one step further and study how quantum-information protocols are to be modified at relativistic speeds and accelerations, as well as in curved spacetimes, an emerging line of research often referred to as relativistic quantum information. That the combination of gravitation and information concepts is fruitful has been famously demonstrated by Bekenstein's discovery of the
entropy of black holes \cite{bekenstein}. This concept was further deepened through the discussions of the black hole information paradox \cite{inf}, and more recently through the so-called AMPS paradox \cite{amps1} ({\it c.f.} Ref. \cite{amps2}), the latter establishing a contradiction between the quantum-information principle of monogamy of entanglement (taken together with the standard theory black hole formation and evaporation) and the equivalence principle.

With superconducting circuits, one can experimentally study standard quantum-information tasks performed with the  usual Alice, Bob, and Eve - this time in relativistic motion. This is enabled by the fact that with quantum circuits one can displace a cavity by using fast-modulated boundary conditions. For example, one can study the degradation of fidelity in quantum teleportation between Alice and Bob due to the nonuniform acceleration of Bob \cite{ivette1}. Since it has been already experimentally proven that two-mode entanglement is generated by the dynamical Casimir effect \cite{pasi}, it is natural to extend this idea to multi-mode cavities. In this case, continuous-variable cluster states can be produced, which can be used as a resource to implement a well-known measurement-based (``one-way'') model of quantum computing \cite{briegel}. Specific protocols for generating for example quadripartite square cluster states,  by using fast changes in the boundary conditions, have been developed \cite{ivette2}.

\subsection{Lattice gauge fields}

The simulation of lattice models for quantum field theory is of great interest since these models have been studied for a long time already on classical computers. The reason for introducing the lattice is that it allows to perform numerical calculations for strongly-coupled theories.

The massless noninteracting Dirac field can be mapped onto the antiferromagnetic XY model of spins by a Jordan-Wigner transformation; in turn, this chain can be simulated by superconducting qubits. However, the difficult and interesting problem that a simulator could solve is the massive and strongly interacting case. One such 1-dimensional model for quantum electrodynamics is due to Schwinger; the model can be discretized to a lattice \cite{kogut}. In this lattice  version, the Hamiltonian is
\begin{eqnarray}
H_{\rm Schwinger} &=&  -J \sum_{l}\left(\psi_{l}^{\dag}S^{+}_{l,l+1}\psi_{l+1}+\psi_{l+1}^{\dag}S^{-}_{l,l+1}\psi_{l}\right) \nonumber \\
& & + m \sum_{l} (-1)^{l} \psi_{l}^{\dag}\psi_{l}
+g \sum_{l}\left(S^{z}_{l,l+1}\right)^2, \label{schw}
\end{eqnarray}
Here $\psi_{l}$ is a fermionic operator (spinless), $S$ is a spin operator, and its $z$ component $S^{z}_{l,l+1}$ describes an electric field between the sites $l$ and $l+1$. The first term in Eq. (\ref{schw}) is kinetic energy, with the hopping of the fermions being accompanied by a flip in the electric field. The second term is a mass (gap) energy term corresponding to the additional energy required to create pairs of particles and antiparticles, and the last term is the analog of the electric energy. One can now apply two standard transformations \cite{zoller}, with the goal of mapping the Hamiltonian Eq. (\ref{schw}) into a form that is readily implementable with superconducting circuit elements. For the fermionic field, one can use the Jordan-Wigner transform
\begin{equation}
\psi_{l} = \exp\left[-i\frac{\pi}{2} \sum_{m<l}(\sigma_{m}^{z}+1)\right],
\end{equation}
and for the  spins $S$ one can employ the Schwinger two-mode representation,
\begin{eqnarray}
S^{z}_{l,l+1} &=& \frac{1}{2}(a^{\dag}_{l}a_{l} - b^{\dag}_{l+1}b_{l+1}); \\
S^{+}_{l,l+1} &=& a^{\dag}_{l}b_{l+1}; \\
S^{-}_{l,l+1} &=& b^{\dag}_{l+1}a_{l}.
\end{eqnarray}
With these transformations, Eq. (\ref{schw}) becomes
\begin{eqnarray}
H_{\rm Schwinger} &=& -J \sum_{l}\left(\sigma_{l}^{\dag}a_{l}^{\dag}b_{l+1}\sigma^{-}_{l+1} + h.c.\right) \nonumber \\ & &
+ \frac{g}{4}\sum_{l}(a_{l}^{\dag}a_{l} - b_{l+1}^{\dag}b_{l+1})^2 + \frac{m}{2}\sum_{l}(-1)^{l} \sigma_{l}^{z}.
\end{eqnarray}
This form of the Hamiltonian is now amenable to simulation by a Josephson circuit designed as a one-dimensional lattice of  qubits with the links between them consisting of two coupled nonlinear oscillators \cite{zoller}.

\section{Simulators of interacting many-body systems}

Interacting many-body problems appear in a variety of contexts in chemistry and solid state physics (quantum phase transitions, superconductivity, quantum magnetism). Below we focus on a few paradigmatic examples that can be solved using superconducting-circuit simulators.

\subsection{Simulation of the Anderson and Kondo models}

The Anderson model describes conduction electrons coupled to an impurity. The starting point for simulating this model is the bosonic Hamiltonian
\begin{eqnarray}
H_{\rm Anderson} &=& -J \sum_{i\geq 1} \left( a_{i+1 \sigma}^{\dag}a_{i\sigma} + h.c.\right)
- t \sum_{\sigma} \left(a^{\dag}_{0\sigma}a_{1\sigma} + h.c. \right) \nonumber\\
& & + E(n_{0\uparrow}, n_{0\downarrow}) + \sum_{i, \sigma}\mu_{i\sigma}n_{i\sigma}.\label{ande}
\end{eqnarray}
This type of Hamiltonian can be realized as a semi-infinite double array of superconducting islands \cite{garciaripoll}, with excess Cooper pair numbers on an island $n_{i\sigma}$, where $\sigma$ is the index of the array and the islands are counted by $i=0,1,2 ...$. Here $J$ is the Josephson energy between junctions everywhere on the array except near the impurity, where it is denoted by $t$; the energy $E(n_{0\uparrow}, n_{0\downarrow})$ is the electrostatic (capacitive) interaction energy between the ``impurity'' islands $i=0$, and $\mu_{i}$ are chemical potentials produced by voltages that have the effect of moving the islands from the degeneracy  points.

Using now a version of the Jordan-Wigner transform that maps bosons to fermions, we can write
\begin{eqnarray}
a_{i\uparrow} &=& c_{i\uparrow}(-1)^{\sum_{j<i}n_{j\uparrow}}, \\
a_{i\downarrow} &=& c_{i\downarrow}(-1)^{\sum_{j<i}n_{j\downarrow}+N_{\uparrow}}, \\
n_{i\sigma} &=& c_{i\sigma}^{\dag}c_{i\sigma} = a_{i\sigma}^{\dag}a_{i\sigma},
\end{eqnarray}
where $c_{i\sigma}$ are fermionic operators and $N_{\sigma}=\sum_{i}n_{i\sigma}$. This allows us to write the
Hamiltonian Eq. (\ref{ande}) of the superconducting circuit simulator described above in a mathematically equivalent form
\begin{eqnarray}
H_{\rm Anderson} &=& -J \sum_{i\geq 1} \left( c_{i+1 \sigma}^{\dag}c_{i\sigma} + h.c.\right)
- t \sum_{\sigma} \left(c^{\dag}_{0\sigma}c_{1\sigma} + h.c. \right) \nonumber \\
& & + E(n_{0\uparrow}, n_{0\downarrow}) + \sum_{i, \sigma}\mu_{i\sigma}n_{i\sigma}.
\end{eqnarray}
This is the well-known Anderson model \cite{Anderson} for hopping electrons with an impurity at $i=0$. The Kondo model, either single-channel or two-channel, can be obtained using similar ideas \cite{garciaripoll}. A different protocol for simulating Anderson localization can be obtained
by implementing a quantum random walk on a lattice with two superconducting qubits per lattice site \cite{ghosh}.

\subsection{Jaynes-Cummings lattices: simulation of  Bose-Hubbard and polaron models}

A straigthforward simulator can be obtained by coupling cavities containing qubits in an array, thus realizing the physics of the Bose-Hubbard model with microwave photons -  but with the additional caveat  of dissipation and thus the necessity of continuous external driving.  Several many-body effects with photons have been proposed in such systems based on photon blockade \cite{hartmann,greentree,brandao}, most notably the realization of a photonic equivalent of the Mott insulator-superfluid transition \cite{angelakis} and the appearance of photon solid phases when the cavity-cavity coupling is nonlinear \cite{fazio}. With superconducting circuits, the standard unit in such a lattice is the transmon-resonator circuit \cite{koch}, where either the resonators or the qubits are coupled.

If the lattice is formed by coupling only the resonators, the Hamiltonian reads
\begin{equation}
H= \sum_{j} H_{j}^{\rm JC} - \sum_{j} J(a^{\dag}_{j}a_{j+1} + a^{\dag}_{j}a_{j+1}),
\end{equation}
where $H_{j}^{\rm JC}$ is the Jaynes-Cummings Hamiltonian of each cavity,
\begin{equation}
H_{j}^{\rm JC} = \hbar\omega_{r} a_{j}^{\dag}a_{j} + \frac{\hbar\omega_{j}}{2}\sigma_{j}^{z}+
\hbar g (a^{\dag}\sigma_{j}^{-} + \sigma_{j}^{+}a),
\end{equation}
and $J$ is the coupling between the resonators.

Recently the dissipation-driven localization transition has been discussed \cite{tureci} and observed \cite{houck} in a Jaynes-Cummings dimer, where only two cavities (left $L$ and right $R$) are used $i=L,R$ .
The dimer effectively realizes a photonic Josephson effect, similar to the oscillations of atoms in a Bose-Einstein condensate in a double well \cite{leggett}. While in the latter case the on-site (well) nonlinearity is provided by the interatomic interaction, for the Jaynes-Cummings dimer this derives from the presence of the qubits and the nonlinearity exists even in the case when they are detuned from their respective cavities due to the ac Stark effect. Other ways of coupling the cavities exist, for example in a ring \cite{girvin} with broken time-reversal symmetry \cite{girvinK}, a scheme resembling a circular array of coupled Bose-Einstein condensates in rotation \cite{paraoanu}.

A well-studied model in many-body physics is the Holstein polaron model, in which the local density of electrons/holes is coupled to the lattice deformation; as a result, polaron excitations are formed due to the dressing of fermionic excitations by (optical) dispersionless phonons. The Holstein model is not analytically solvable, although approximation techniques giving results in good agreement with Monte Carlo simulation do exist, for example the Toyozawa ansatz. Thus, a Holstein simulator would be a useful tool. The elementary unit of such a device can be the Jaynes-Cummings cavity-qubit system, as used in the standard transmon circuit, with the additional feature that the qubits are coupled via a tunable interaction Hamiltonian \cite{mei}. The total Hamiltonian of the system is therefore
\begin{eqnarray}
H&=& \sum_{n}\left[\hbar\omega_{z}a^{\dag}_{n}a_{n}+\frac{\hbar\omega_{c}}{2}\sigma_{n}^{z} + \hbar g (a_{n}^{\dag}\sigma_{n}^{-} + \sigma_{n}^{+}a_{n}) \right] + \nonumber \\
& & + \sum_{n} \left[-t_{0}(\sigma_{n}^{+}\sigma_{n+1}^{-} + \sigma_{n+1}^{+}\sigma_{n}^{-}) + 2\epsilon_{0} \cos (\omega_{d} t) (a_{n}+ a_{n}^{\dag} )\right]. \label{htot}
\end{eqnarray}
The frequencies of the resonators are denoted here by $\omega_{r}$, that of the qubits by $\omega_{c}$, the coupling between the qubits and the corresponding resonator is $g$, the driving of the resonators is done by a field with frequency $\omega_{d}$ coupled to each resonator by $\epsilon_{0}$,
and the qubit-qubit interaction is characterized by a hopping matrix element $t_{0}$, which can be tuned by applying an external magnetic field. In the dispersive regime (defined as $|\Delta | \gg g$, where the detuning  is $\Delta = \omega_{c}-\omega_{z}$) we can apply a Schrieffer-Wolff transformation $U_{n} = \exp \left[-\frac{g}{\Delta} (\sigma_{n}^{+}a_{n}- a_{n}^{\dag}\sigma_{n}^{-})\right]$
to each of the Jaynes-Cummings units, resulting in the elimination of the qubit-resonator coupling; this is followed by a displacement transformation $a_{n}  \rightarrow a_{n}- \epsilon_{0}/\hbar\delta\omega$, where $\delta\omega = \omega_{c}+ \chi - \omega_{d}$ and $\chi = g^{2}/\Delta$ is the Stark shift. As a result, in the interaction picture and under the assumption $\epsilon_{0} \gg \hbar \delta\omega$ (which allows to neglect the vacuum term $-\hbar \chi (\sigma_{n}^{z}+1 )a^{\dag}_{n}a_{n}$) the Hamiltonian Eq. (\ref{htot}) becomes \cite{mei}
\begin{equation}
H_{\rm eff} = \sum_{n} \hbar \delta\omega\left[a_{n}^{\dag}a_{n} + g_{H}\frac{\sigma_{n}^{z} + 1}{2}(a_{n}+ a_{n}^{\dag})\right]
 -\sum_{n} t_{0}(\sigma_{n}^{+}\sigma_{n+1}^{-} + \sigma_{n+1}^{+}\sigma_{n}^{-}),
\end{equation}
where $g_{H}\delta\omega = 2\epsilon_{0}\chi /\hbar\delta\omega$. To obtain from this Hamiltonian the one-dimensional Holstein model \cite{Holstein}, we can represent the qubit operators as Fermi fields via a Jordan-Wigner transform,
\begin{eqnarray}
\sigma_{n}^{+} &=& c_{n}^{\dag}\prod_{m=1}^{n-1}e^{i \pi c_{m}^{\dag}c_{m}}, \\
\sigma_{n}^{z} &=& 2c_{n}^{\dag}c_{n}-1,
\end{eqnarray}
resulting in
\begin{equation}
H_{\rm Holstein} = \sum_{n}\left[\hbar\delta\omega a^{\dag}_{n}a_{n} + \hbar\delta\omega g_{H}c^{\dag}_{n}c_{n}(a^{\dag}_{n} + a_{n}) - t_{0} (c_{n}^{\dag}c_{n+1}+c_{n+1}^{\dag}c_{n})\right],
\end{equation}
where now the first term plays the role of the phonon Hamiltonian, the second describes the local density-coupling of the fermions to the phonons, and the last term is the fermionic hopping term. The form of the Hamiltonian above allows the independent variation of each of the parameters entering the Hamiltonian. Indeed, $\delta\omega$ can be changed by changing the frequency of the driving field or the detuning $\Delta$; the dimensionless Holstein coupling $g_{H}$ can be modified via the amplitude of the driving field, which is directly proportional to $\epsilon_{0}$; and the hopping matrix term $t_{0}$ can be controlled by using a SQUID-type configuration for the coupling and applying an external magnetic field.  Thus, every parameter regime of this model, including that of formation of ``small'' polarons \cite{Holstein}, can be reached.

The modification of the Holstein model to the case when the phonons are coupled to the hopping matrix element instead of the density is called the SSH (Su-Schrieffer-Heeger) model \cite{ssh}. In this case the momentum of the electrons as they move between lattice sites is modulated by the displacement of the lattice $u_{n+1}-u_{n}$, where $u_{n}\approx a^{\dag}_{n} + a_{n}$, leading to coupling Hamiltonians of the type
\begin{equation}
H_{\rm SSH} = \hbar\delta\omega g_{SSH}(c^{\dag}_{n}+c_{n+1})(a^{\dag}_{n+1} + a_{n+1}-a^{\dag}_{n} - a_{n+1}).
\end{equation}
This type of coupling appears in a variety of
transport models (especially by nonlinear excitations such as solitons) of $\pi$-electron systems,
such as the  polyacetylene molecule and other organic semiconductors, as well as carbon nanotubes and graphene nanostructures.

Interestingly, the SSH model evades the conditions of validity of the Gerlach-L\o wen theorem, which asserts that for momentum-independent couplings models - such as the Holstein model - the ground state is smooth in the coupling parameters, thus these models do not have phase transitions. The phases of the SSH-coupled model are subject to intensive theoretical investigations, and sharp transitions as a function of coupling have been predicted recently \cite{ssh1,ssh2,ssh3,ssh4}. An experimental validation of these predictions would be very valuable. A quantum simulator for the SSH model using superconducting circuits can be realized along the same lines as the Holstein simulator described above \cite{tian}. To realize the modulation of the electron momentum, one has to engineer a tunable coupling between the qubits using a SQUID loop into which the flux of the resonators would couple \cite{tian}.

\subsection{Spin lattices and adiabatic simulators}

A special case of quantum simulators is that of adiabatic machines \cite{zagoskin}; they solve a specific problem, namely that of optimization by quantum annealing. This direction has been consistently pursued by D-wave, a company that has designed and operated processors comprising 128 (first-generation) and 512 (second-generation) flux qubits with tunable couplings.

The Hamiltonian of a typical D-wave device can be written in the form \cite{geordie}
\begin{equation}
H (t) = \Gamma (t) \sum_{i=1}^{N}\Delta_{i}\sigma_{i}^{x}+\Lambda(t)H_{\rm P},
\end{equation}
where $\Delta_i$ is the gap between states with clockwise and counterclockwise persistent currents, and the Hamiltonian $H_{\rm P}$ has the Ising-type form
\begin{equation}
H_{\rm P} = \sum_{i=1}^{N}h_{i}\sigma_{i}^{z} + \sum_{i,j=1}^{N}J_{ij}\sigma_{i}^{z}\sigma_{j}^{z}.
\end{equation}
Quantum annealing proceeds by adiabatically decreasing $\Gamma (t)$ from 1 to 0 and increasing $\Lambda$ from 0 to 1 monotonically in time, such that the system ends up in the ground state of $H_{\rm P}$. The fact that annealing is realized by quantum-mechanical tunneling and not just thermal activation has been tested so far for 8 qubits \cite{geordie}, though in more recent D-wave devices, the number of qubits that stay quantum-coherent could be larger.

Many well-known difficult problems from various fields (pattern matching, traveling salesman, spin glasses, sampling problems, efficient data compression, {\it etc.}) can be mathematically mapped onto the adiabatic quantum protocol. Very recently, finding the ground state of the Miyazawa-Jernigan model of protein folding has been achieved on this processor \cite{protein}. Protein folding is very important in biology and medicine - incorrect protein folding causes for example the Parkinson and the Alzheimer diseases.

The issue of  the  ``quantumness'' of the D-wave processor has attracted recently a lot of interest. Several comparison tests have been performed to see if the claim that the calculation is faster than what is achievable with classical means stands up. So far, the D-wave computer has been shown to be more efficient when tested against general-purpose software, but not against optimized classical codes. The issue of the scale over which quantum coherence is maintained in this processor is intensely investigated \cite{troyer}, and clear benchmarking criteria for quantum speedup have been proposed \cite{troyerlidar}. Quantum error correction would certainly improve the performance of these machines. Recently, an error-correction protocol for quantum annealing has been demonstrated on 344 qubits \cite{lidar}.

\subsection{The Tavis-Cummings model}

The  Tavis-Cummings model \cite{Tavis} is a model originally proposed to describe the interaction between an ensemble of noninteracting molecules and the electromagnetic field. It is defined by the Hamiltonian
\begin{equation}
H_{\rm TC} = \hbar\omega_{r} a^{\dag}a+ \sum_{j=1}^{N}\left[\frac{\hbar\omega_{j}}{2}\sigma_{j}^{z}+
\hbar g (a^{\dag}\sigma_{j} + \sigma_{j}^{+}a)\right],
\end{equation}
where $\omega_{r}$ is the frequency of the resonator, $\omega_{j}$ is that of each of the $N$ qubits, and $g$ is the coupling between the qubit and the resonator. The interesting feature of the Tavis-Cummings model is the formation of collective multiqubit states with a $\sqrt{N}$ collective dipole strength. This $\sqrt{N}$ scaling has been verified for $N=3$ transmon qubits placed in a coplanar waveguide resonator \cite{fink}. In this case the transitions visible at degeneracy $\omega_{j}=\omega_{r}$ are those between the ground state $|\rm g,\rm g,\rm g\rangle \otimes |0\rangle$ and two ``bright'' states containing one excitation, either in the resonator or distributed among the the qubits,
\begin{equation}
|3, 1\pm\rangle = \frac{1}{\sqrt{2}}\left[|\rm g,\rm g,\rm g\rangle \otimes |1\rangle \pm |W_{3}\rangle \otimes |0\rangle \right],
\end{equation}
where $|W_{3}\rangle$ is the 3-qubit $W$-state defined as
\begin{equation}
|W_{3}\rangle = \frac{1}{\sqrt{3}}\left(|\rm e,\rm g,\rm g\rangle +|\rm g,\rm e,\rm g\rangle + |\rm g,\rm g,\rm e\rangle\right).
\end{equation}
The frequency separation between the transitions $|\rm g,\rm g,\rm g\rangle \otimes |0\rangle \rightarrow |3, 1\pm\rangle $ is $\sqrt{3}g/\pi$, allowing to verify directly the $\sqrt{3}$ scaling from spectroscopic measurements \cite{fink}. $W$ states, in which the excitation is coherently shared between a large number of qubits, are interesting from a fundamental point of view, since it has been shown that they lead to
direct logical contradictions with local realism \cite{vedral1,vedral2}.
To increase the number of qubits one needs to have a qubit smaller in size than the transmon, so they can fit along the resonator. Recently, such a sample with 20 flux qubits inserted in a resonator has been fabricated and studied \cite{ustinov}, allowing to check the $\sqrt{N}$ scaling and the formation of collective modes for up to $N=8$.

\section{Conclusions}

The technology of superconducting circuits is at the forefront of a research direction attempting to realize quantum simulators. Such devices would be tremendously interesting for scientific research, and would have a high impact on solving certain mathematical, physical, and computational problems. Simulating the behavior of real systems faster than what is possible with classical computers would be a turning point in computational science in general. Also, using quantum simulators one could in principle have access to a wider range of parameters to explore, opening the way to the discovery of novel effects and states of matter.


\begin{thebibliography}{}

\bibitem{Feynman} R. Feynman, Simulating physics with computers, Int. J. Theor. Phys. {\bf 21}, 467–488 (1982).

\bibitem{deutsch} D. Deutsch, Quantum Theory, the Church-Turing Principle and the Universal Quantum Computer,
Proc. R. Soc. London A {\bf 400}, 97 (1985).

\bibitem{lloyd} S. Lloyd, Universal Quantum Simulators, Science {\bf 273}, 1073 (1996).

\bibitem{nielsen} M. A. Nielsen and I. C. Chuang, Quantum Computation and Quantum Information, Cambridge University Press, Cambridge (2000).

\bibitem{physperspect} G. S. Paraoanu, Quantum computing: theoretical possibility versus practical possibility, Phys. Perspect. {\bf 13}, 359 (2011).

\bibitem{buluta1} I. Buluta, and F. Nori, Quantum simulators, Science {\bf 326}, 108 (2009).

\bibitem{buluta2} I. Buluta. S. Ashhab, and F. Nori, Natural and artificial atoms for quantum computation,  Rep. Prog. Phys. {\bf 74}, 104401 (2011).

\bibitem{ciraczoller} J. I. Cirac and P. Zoller, Goals and opportunities in quantum simulation, Nature Physics {\bf 8}, 264 (2012).

\bibitem{georgescu} I. M. Georgescu, S. Ashhab, and F. Nori, Quantum Simulation, Rev. Mod. Phys. {\bf 86}, 153 (2014).

\bibitem{lewenstein} P. Hauke, F. M. Cucchietti, L. Tagliacozzo, I. Deutsch, and M. Lewenstein,
Can one trust quantum simulators?, Rep. Prog. Phys. {\bf 75}, 082401 (2012).

\bibitem{lewensteinM} M. Lewenstein, {\it et al.} Ultracold atomic gases in optical lattices: Mimicking
condensed matter physics and beyond, Adv. Phys. {\bf 56}, 243 (2007).

\bibitem{shnirman} Y. Makhlin, G. Sch\"on, and A. Shnirman, Quantum-state engineering with Josephson-junction devices, Rev. Mod. Phys. {\bf 73}, 357 (2001).

\bibitem{devoret} M. H. Devoret, A. Wallraff, J. M. Martinis, Superconducting Qubits: A Short Review, arXiv:cond-mat/0411174.

\bibitem{clarke} J. Clarke and F. Wilhelm, Superconducting quantum bits, Nature {\bf 453}, 1031 (2008).

\bibitem{wendin} G. Wendin and V.S. Shumeiko, Superconducting Quantum Circuits, Qubits and Computing, in ``Handbook of Theoretical and Computational Nanotechnology'' (eds. M. Rieth and W. Schommers)  {\bf 3}, pp. 223-309, American Scientific Publishers, Los Angeles (2006).

\bibitem{girvinrev} R. J. Schoelkopf and S. M. Girvin, Wiring up quantum systems, Nature {\bf 451}, 664 (2008).

\bibitem{nori} J. Q. You and F. Nori, Atomic physics and quantum optics using superconducting circuits, Nature {\bf 474}, 589 (2011).

\bibitem{schoelkopf} M. H. Devoret, and R. J. Schoelkopf, Superconducting circuits for quantum information: an outlook, Science {\bf 339}, 1169 (2013).


\bibitem{cnotmartinis} R. C. Bialczak, M. Ansmann, M. Hofheinz, E. Lucero, M. Neeley, A. D. O'Connell, D. Sank, H. Wang, J. Wenner, M. Steffen, A. N. Cleland, and J. M. Martinis, Quantum process tomography of a universal entangling gate implemented with Josephson phase qubits, Nature Physics {\bf 6}, 409 (2010).

\bibitem{singleshotesteve} A. Dewes, F. R. Ong, V. Schmitt, R. Lauro, N. Boulant, P. Bertet, D. Vion, and D. Esteve, Characterization of a Two-Transmon Processor with Individual Single-Shot Qubit Readout, Phys. Rev. Lett. {\bf 108}, 057002 (2012).

\bibitem{niskanen} A. O. Niskanen, K. Harrabi, F. Yoshihara, Y. Nakamura, S. Lloyd, and J. S. Tsai, Quantum Coherent Tunable Coupling of Superconducting Qubits, Science {\bf 316}, 723 (2007).


\bibitem{majer} J. Majer, J. M. Chow, J. M. Gambetta, J. Koch, B. R. Johnson, J. A. Schreier, L. Frunzio, D. I. Schuster, A. A. Houck, A. Wallraff, A. Blais, M. H. Devoret, S. M. Girvin, and R. J. Schoelkopf, Coupling superconducting qubits via a cavity bus, Nature {\bf 449}, 443-447 (2007).


\bibitem{coherentcoupling}
 R. C. Bialczak, M. Ansmann, M. Hofheinz, M. Lenander, E. Lucero, M. Neeley, A. D. O'Connell, D. Sank, H. Wang, M. Weides, J. Wenner, T. Yamamoto, A. N. Cleland, and J. M. Martinis,
Quantum Coherent Tunable Coupling
of Superconducting Qubits, Phys. Rev. Lett. {\bf 106}, 060501 (2011).


\bibitem{mw1}  C. Rigetti, A. Blais, and M. Devoret,
Protocol for universal gates in optimally biased superconducting qubits, Phys. Rev. Lett. {\bf 94}, 240502 (2005).

\bibitem{mw2} G. S. Paraoanu, Microwave-induced coupling of superconducting qubits, Phys. Rev. B {\bf 74}, 140504(R) (2006).

\bibitem{mw3} J. Li, K. Chalapat, and G. S. Paraoanu, Entanglement of superconducting qubits via
microwave fields: classical and quantum regimes, Phys. Rev. B {\bf 78}, 064503 (2008).

\bibitem{mw4} C. Rigetti, and M. Devoret, Fully microwave-tunable universal gates in superconducting qubits with linear couplings and fixed transition frequencies, Phys. Rev. B  {\bf 81}, 134507 (2010).

\bibitem{mw5} J. M. Chow, J. M. Gambetta, A. W. Cross, S. T. Merkel, C. Rigetti and M. Steffen, Microwave-activated conditional-phase gate for superconducting qubits, New J. Phys. {\bf 15}, 115012 (2013).

\bibitem{sw1} D. Vion, A. Aassime, A. Cottet, P. Joyez, H. Pothier, C. Urbina, D. Esteve, and M. H. Devoret, Manipulating the Quantum State of an Electrical Circuit, Science {\bf 296}, 886 (2002).

\bibitem{sw2} I. Chiorescu, Y. Nakamura, C. J. P. M. Harmans, and J. E. Mooij, Coherent Quantum Dynamics of a Superconducting Flux Qubit, Science {\bf 299}, 1869 (2003).

\bibitem{sw3} J. Claudon, F. Balestro, F. W. J. Hekking, and O. Buisson, Coherent Oscillations in a Superconducting Multilevel Quantum System, Phys. Rev. Lett. {\bf 93}, 187003 (2004).

\bibitem{sw4} J. M. Martinis, S. Nam, J. Aumentado, and C. Urbina, Rabi Oscillations in a Large Josephson-Junction Qubit, Phys. Rev. Lett. {\bf 89}, 117901 (2002).

\bibitem{sw5} G. S. Paraoanu, Running-phase state in a Josephson washboard potential, Phys. Rev. B {\bf 72}, 134528 (2005).

\bibitem{sw6} J. Walter, E. Thol\'en, D. B. Haviland, and J. S\"ostrand, Pulse and hold strategy for switching current measurements, Phys. Rev. B {\bf 75}, 094515 (2007).

\bibitem{transmon1} A. Wallraff, D. I. Schuster, A. Blais, L. Frunzio, R.-S. Huang, J. Majer, S. Kumar, S. M. Girvin, and R. J. Schoelkopf, Strong Coupling of a Single Photon to a Superconducting Qubit Using Circuit Quantum Electrodynamics, Nature {\bf 431}, 162 (2004);

\bibitem{transmon2} A. Blais, R.-S. Huang, A. Wallraff, S. M. Girvin, and R. J. Schoelkopf, Cavity quantum electrodynamics for superconducting electrical circuits: An architecture for quantum computation, Phys. Rev. A {\bf 69}, 062320 (2004).


\bibitem{partial1} N. Katz, M. Ansmann, R. C. Bialczak, E. Lucero, R. McDermott, M. Neeley, M. Steffen, E. M. Weig, A. N. Cleland, John M. Martinis, and A. N. Korotkov, Coherent State Evolution in a Superconducting Qubit from Partial-Collapse Measurement, Science {\bf 312}, 1498 (2006).

\bibitem{partial2} G. S. Paraoanu, Interaction-free measurements with superconducting qubits, Phys. Rev. Lett. {\bf 97}, 180406 (2006).

\bibitem{partial3} N. Katz, M. Neeley, M. Ansmann, R. C. Bialczak, M. Hofheinz, E. Lucero, A. O'Connell, H. Wang, A. N. Cleland, J. M. Martinis, and A. N. Korotkov, Reversal of the Weak Measurement of a Quantum State in a Superconducting Phase Qubit, Phys. Rev. Lett. {\bf 101}, 200401 (2008).

\bibitem{partial4}  A. N. Jordan, and A.  N. Korotkov,  Uncollapsing the wavefunction by undoing quantum measurements, Contemp. Phys. {\bf 51}, 125 (2010).

\bibitem{partial5} G. S. Paraoanu, Generalized partial measurements, Europhys. Lett. {\bf 93}, 64002 (2011).

\bibitem{partial6} G. S. Paraoanu, Partial measurements and the realization of quantum-mechanical counterfactuals, Found. Phys. {\bf 41}, 1214 (2011).

\bibitem{singleshot} M. D. Reed, L. DiCarlo, B. R. Johnson, L. Sun, D. I. Schuster, L. Frunzio, and R. J. Schoelkopf, High-Fidelity Readout in Circuit Quantum Electrodynamics Using the Jaynes-Cummings Nonlinearity,
    Phys. Rev. Lett. {\bf 105}, 173601 (2010).

\bibitem{bif}  F. Mallet, F. R. Ong, A. Palacios-Laloy, F. Nguyen, P. Bertet, D. Vion, D. Esteve, Single-shot qubit readout in circuit quantum electrodynamics, Nature Physics {\bf 5} 791 (2009).

\bibitem{feedback1} D. Rist\'e, C. C. Bultink, K. W. Lehnert, and L. DiCarlo, Feedback control of a solid-state qubit using high-fidelity projective measurement, Phys. Rev. Lett. {\bf 109}, 240502 (2012).

\bibitem{feedback2} R. Vijay, C. Macklin, D. H. Slichter, S. J. Weber, K. W. Murch, R. Naik, A. N. Korotkov, I. Siddiqi, Quantum feedback control of a superconducting qubit: Persistent Rabi oscillations, Nature {\bf 490}, 77 (2012).

\bibitem{feedback3} P. Campagne-Ibarcq, E. Flurin, N. Roch, D. Darson, P. Morfin, M. Mirrahimi, M. H. Devoret, F. Mallet, and B. Huard, Persistent Control of a Superconducting Qubit by Stroboscopic Measurement Feedback, Phys. Rev. X {\bf 3}, 021008 (2013).

\bibitem{feedback4} G. de Lange, D. Rist\'e, M. J. Tiggelman, C. Eichler, L. Tornberg, G. Johansson, A. Wallraff, R. N. Schouten, L. DiCarlo, Reversing quantum trajectories with analog feedback, arXiv:1311.5472.

\bibitem{zagoskinmetamaterials1} A. L. Rakhmanov, A. M. Zagoskin, Sergey Savel'ev, Franco Nori, Quantum metamaterials: Electromagnetic waves in a Josephson qubit line, Phys. Rev. B {\bf 77}, 144507 (2008).

\bibitem{zagoskinmetamaterials2} A. M. Zagoskin, Quantum Engineering: Theory and Design of Quantum Coherent Structures, Cambridge University Press, Cambridge (2011), Ch. 6.1.

\bibitem{adiabatic} A. Das and B. K. Chakrabarti, Quantum annealing and analog quantum computation, Rev. Mod. Phys. {\bf 80}, 1061 (2008).

\bibitem{trotter1} H. F. Trotter, On the product of semi-groups of operators,  Proc. Amer. Math. Soc. {\bf 10}, 545 (1959).

\bibitem{trotter2} M. Suzuki, Decomposition formulas of exponential operators and Lie exponentials with some applications to quantum mechanics and statistical physics, J. Math. Phys. {\bf 26}, 601 (1985).

\bibitem{gates1} M. Neeley, R. C. Bialczak, M. Lenander, E. Lucero, M.
Mariantoni, A. D. O'Connell, D. Sank, H. Wang, M. Weides,
J. Wenner, Y. Yin, T. Yamamoto, A. N. Cleland and
J. M. Martinis, Generation of three-qubit entangled states using superconducting phase qubits,
Nature {\bf 467}, 570 (2010).

\bibitem{gates2} A. Fedorov, L. Steffen, M. Baur, M. P. da Silva, and A.
Wallraff, Implementation of a Toffoli gate with superconducting circuits, Nature {\bf 481}, 170 (2012).

\bibitem{gates3} J. M. Chow, J. M. Gambetta, A. D. C\'orcoles, S. T.
Merkel, J. A. Smolin, C. Rigetti, S. Poletto, G. A. Keefe,
M. B. Rothwell, J. R. Rozen, M. B. Ketchen, and M.
Steffen, Universal Quantum Gate Set Approaching Fault-Tolerant Thresholds with Superconducting Qubits,
Phys. Rev. Lett. {\bf 109}, 060501 (2012).

\bibitem{gates4}
M. D. Reed, L. DiCarlo, S. E. Nigg, L. Sun, L. Frunzio, S.
M. Girvin and R. J. Schoelkopf, Realization of Three-Qubit Quantum Error Correction with Superconducting Circuits, Nature {\bf 482}, 382 (2012).


\bibitem{wallraff} U. Las Heras, A. Mezzacapo, L. Lamata, S. Filipp, A. Wallraff, E. Solano, Digital Quantum Simulation of Spin Systems in Superconducting Circuits, arXiv:1311.7626.

\bibitem{chuang} K. R. Brown, R. J. Clark, and I. L. Chuang, Limitations of Quantum Simulation Examined by Simulating a Pairing Hamiltonian Using Nuclear Magnetic Resonance, Phys.
Rev. Lett. {\bf 97}, 050504 (2006).

\bibitem{AT1} M. A. Sillanp\"a\"a, J. Li, K. Cicak, F. Altomare, J. I. Park, R. W. Simmonds,
G. S. Paraoanu, and P. J. Hakonen, Autler-Townes effect in a superconducting three-level system,
Phys. Rev. Lett. {\bf 103}, 193601 (2009).

\bibitem{AT2} M. Baur, S. Filipp, R. Bianchetti, J. M. Fink, M. G\"oppl, L. Steffen, P. J. Leek, A. Blais, and A. Wallraff, Measurement of Autler-Townes and Mollow Transitions in a Strongly Driven Superconducting Qubit, Phys. Rev. Lett. {\bf 102}, 243602 (2009).

\bibitem{AT3} A. A. Abdumalikov Jr., O. Astafiev, A. M. Zagoskin, Yu. A. Pashkin, Y. Nakamura, and J. S. Tsai, Electromagnetically Induced Transparency on a Single Artificial Atom, Phys. Rev. Lett. {\bf 104}, 193601 (2010).

\bibitem{AT4} W. R. Kelly, Z. Dutton, J. Schlafer, B. Mookerji, T. A. Ohki, J. S. Kline, and D. P. Pappas, Direct Observation of Coherent Population Trapping in a Superconducting Artificial Atom, Phys. Rev. Lett. {\bf 104}, 163601 (2010).

\bibitem{Neely} M. Neeley, M. Ansmann, R. C. Bialczak, M. Hofheinz, E. Lucero, A. D. O'Connell, D. Sank, H. Wang, J. Wenner, A. N. Cleland, M. R. Geller, J. M. Martinis, Emulation of a quantum spin with a superconducting phase qudit, Science {\bf 325}, 722–725 (2009).

\bibitem{jian_prb} J. Li, G. S. Paraoanu, K. Cicak, F. Altomare, J. I. Park, R. W. Simmonds, M. A. Sillanp\"a\"a, and
P. J. Hakonen, Decoherence, Autler-Townes effect, and dark states in two-tone driving of a three-level
superconducting system, Phys. Rev. B. {\bf 84}, 104527 (2011).

\bibitem{scirep} J. Li, G. S. Paraoanu, K. Cicak, F. Altomare, J. I. Park, R.
W. Simmonds, M. A. Sillanp\"a\"a and P. J. Hakonen,
Dynamical Autler-Townes control of a phase qubit, Sci. Rep. {\bf 2}, 645 (2012).

\bibitem{motion} J. Li, M. P. Silveri, K. S. Kumar, J. M. Pirkkalainen, A. Veps\"al\"ainen, W. C. Chien, J. Tuorila, M. A. Sillanp\"a\"a, P. J. Hakonen, E. V. Thuneberg, and G. S. Paraoanu,
Motional averaging in a superconducting qubit, Nat. Commun. {\bf 4}, 1420 (2013).

\bibitem{zitt} J. S. Pedernales, R. Di Candia, D. Ballester, E. Solano, Quantum Simulations of Relativistic Quantum Physics in Circuit QED, New J. Phys. {\bf 15}, 055008 (2013).

\bibitem{discrete} B. Doucot, L. B. Ioffe, and J. Vidal, Discrete non-Abelian gauge theories in Josephson-junction arrays and quantum computation, Phys. Rev. B {\bf 69}, 214501 (2004).

\bibitem{preskill} S. P. Jordan, K. S. M. Lee, J. Preskill, Quantum Algorithms for Quantum
Field Theories, Science {\bf 336}, 1130 (2013).

\bibitem{pasi} P. L\"ahteenm\"aki, G. S. Paraoanu, J. Hassel, and P. J. Hakonen, Dynamical Casimir effect in
a Josephson metamaterial, Proc. Natl. Acad. Sci. U.S.A. {\bf 110}, 4234-4238 (2013).

\bibitem{wilson} C. M. Wilson, G. Johansson, A. Pourkabirian, M. Simoen, J. R. Johansson, T. Duty, F. Nori, and
P. Delsing, Observation of the dynamical Casimir effect in a superconducting circuit, Nature {\bf 479}, 376-379 (2011).

\bibitem{early1} V. V. Dodonov, V. I. Man'ko, and O. V. Man'ko, Correlated states in quantum electronics (resonant circuit), J. Sov. Laser Res. {\bf 10}, 413 (1989).

\bibitem{early2} E. Yablonovitch, J. P. Heritage, D. E. Aspnes, and Y. Yafet, Virtual Photoconductivity, Phys. Rev. Lett. {\bf 63}, 976 (1989).

\bibitem{early3} E. Yablonovitch, Accelerating reference frame for electromagnetic waves in a rapidly growing plasma: Unruh-Davies-Fulling-DeWitt radiation and the nonadiabatic Casimir effect, Phys. Rev. Lett. {\bf 62}, 1742 (1989).

\bibitem{enta1} J. Li and  G. S. Paraoanu, Generation and propagation of entanglement in driven coupled-qubit systems, New J. Phys. {\bf 11}, 113020 (2009).

\bibitem{enta2} S. Felicetti, M. Sanz, L. Lamata, G. Romero, G. Johansson, P. Delsing, and E. Solano, Dynamical Casimir effect entangles artificial atoms, arXiv:1402.4451.

\bibitem{zeilinger}  T. Fujii, S.  Matsuo, N. Hatakenaka, S.  Kurihara, and A. Zeilinger, Quantum circuit analog of the dynamical Casimir effect, Phys. Rev. B {\bf 84}, 174521 (2011).

\bibitem{nation} P. D. Nation, M. P. Blencowe, A. J. Rimberg, E. Buks,  Analogue Hawking Radiation in a dc-SQUID Array Transmission Line, Phys. Rev. Lett. {\bf 103}, 087004  (2009).

\bibitem{viser} C. Barcel\'o, S. Liberati, and M. Viser, Analogue Gravity, Living Rev. Rel. {\bf 14}, 3 (2011)

\bibitem{volovik} G. E. Volovik, The Universe in a Helium Droplet  (Clarendon Press, Oxford, 2003).

\bibitem{nationrmp} P. D. Nation, J. R. Johansson, M. P. Blencowe, and F. Nori, Stimulating uncertainty: Amplifying the quantum vacuum with superconducting circuits,
Rev. Mod. Phys. {\bf 84}, 1 (2012).

\bibitem{bekenstein}J. D. Bekenstein, Black Holes and Entropy, Phys. Rev. D. {\bf 7}, 2333 (1973).

\bibitem{inf}  S. W. Hawking, Breakdown of Predictability in Gravitational Collapse, Phys. Rev. D
{\bf 14}, 2460 (1976).

\bibitem{amps1} A. Almheiri, D. Marolf, J. Polchinski, and J. Sully,  Black Holes: Complementarity or Firewalls?, J. High Energy Phys. {\bf 02}, 062 (2013).

\bibitem{amps2} S. L. Braunstein, Black hole entropy as entropy of
entanglement, or it's curtains for the equivalence principle,
[arXiv:0907.1190v1] published as S. L. Braunstein, S.
Pirandola and K. \.Zyczkowski, Better Late than Never: Information
Retrieval from Black Holes, Physical Review Letters {\bf 110}, 101301
(2013).

\bibitem{ivette1} N. Friis, A. R. Lee, K. Truong, C. Sab\'in, E. Solano, G\"oran Johansson, and I. Fuentes, Relativistic Quantum Teleportation with Superconducting Circuits, Phys. Rev. Lett. {\bf 110}, 113602 (2013).

\bibitem{briegel} H. J. Briegel, D. E. Browne, W. D\"ur, R. Raussendorf, and M. Van den Nest,
Measurement-based quantum computation, Nature Physics {\bf 5}, 19 (2009).

\bibitem{ivette2} D. E. Bruschi, C. Sab\'in, P. Kok, G\"oran Johansson, Per Delsing, and Ivette Fuentes, Towards universal quantum computation through relativistic motion, arXiv:1311.5619.

\bibitem{kogut} T. Banks, L. Susskind, and J. Kogut, Strong-coupling calculations of lattice gauge theories: (1 + 1)-dimensional exercises, Phys. Rev. D {\bf 13}, 1043–1053 (1976).

\bibitem{zoller} D. Marcos, P. Rabl, E. Rico, P. Zoller, Superconducting Circuits for Quantum Simulation of Dynamical Gauge Fields, Phys. Rev. Lett. {\bf 111}, 110504 (2013).

\bibitem{Anderson} P. W. Anderson, Localized Magnetic States in Metals,
Phys. Rev. {\bf 124}, 41 (1961).

\bibitem{garciaripoll}J. J. Garc\'ia-Ripoll, E. Solano, and M. A. Martin-Delgado, Quantum simulation of Anderson and Kondo lattices with superconducting qubits, Phys. Rev. B {\bf 77}, 024522 (2008).

\bibitem{ghosh} J. Ghosh, Simulating Anderson localization via a quantum walk on a one-dimensional lattice
of superconducting qubits, Phys. Rev. A {\bf 89}, 022309 (2014).

\bibitem{hartmann} M. J. Hartmann, Strongly interacting polaritons in coupled
arrays of cavities, Nature Physics {\bf 2}, 849 (2006).

\bibitem{greentree} A. Greentree, C. Tahan, J. Cole, and L. Hollenberg,
Quantum phase transitions of light, Nature Physics {\bf 2}, 856 (2006).

\bibitem{brandao} M. J. Hartmann, F. G. S. L. Branda\~o, and M. B. Plenio, Quantum Many-Body Phenomena in Coupled Cavity
Arrays, Laser \& Photon. Rev. {\bf 2}, 527 (2008).

\bibitem{angelakis} D. G. Angelakis, M. F. Santos, S. Bose, Photon blockade induced Mott transitions and XY spin models in coupled cavity arrays, Phys. Rev. A {\bf 76}, 031805 (2007).

\bibitem{fazio}
J. Jin, D. Rossini, R. Fazio, M. Leib, M. J. Hartmann, Photon solid phases in driven arrays of nonlinearly coupled cavities, Phys. Rev. Lett. {\bf 110}, 163605 (2013).

\bibitem{koch} S. Schmidt and J. Koch, Circuit QED lattices: towards quantum simulation with superconducting circuits, Annalen der Physik {\bf 525}, 395-412 (2013).

\bibitem{tureci}
S. Schmidt, D. Gerace, A. A. Houck, G. Blatter, and H. E. T\"ureci, Nonequilibrium delocalization-localization transition of photons in circuit quantum electrodynamics, Phys. Rev. B {\bf 82}, 100507 (2010).

\bibitem{houck} J. Raftery, D. Sadri, S. Schmidt, H. E.T\"ureci,
and A. A. Houck, Observation of a Dissipation-Induced Classical to Quantum Transition,
arXiv:1312.2963.

\bibitem{leggett} Gh.-S. Paraoanu, S. Kohler, F. Sols, and A. J. Leggett, The
Josephson plasmon as a Bogoliubov quasiparticle, J. Phys.
B {\bf 34}, 4689 (2001).

\bibitem{girvin} A. Nunnenkamp, Jens Koch, and S. M. Girvin, Synthetic gauge fields and homodyne transmission in Jaynes-Cummings lattices, New J. Phys. {\bf 13}, 095008 (2011).

\bibitem{girvinK}  J. Koch, A. A. Houck, K. Le Hur, and S. M. Girvin, Time-reversal symmetry breaking in circuit-QED based photon lattices, Phys. Rev. A {\bf 82}, 043811 (2010).

\bibitem{paraoanu} Gh.-S. Paraoanu, Persistent currents in a circular array of Bose-Einstein condensates,
Phys. Rev. A {\bf 67}, 023607 (2003)

\bibitem{mei} F. Mei, V. M. Stojanovi\'c, I. Siddiqi, and L. Tian, Analog Superconducting Quantum Simulator for Holstein Polarons, Phys. Rev. B {\bf 88}, 224502 (2013).

\bibitem{Holstein} T. Holstein, Studies of polaron motion: Part II. The ``small'' polaron, Ann. Phys. (N.Y.) {\bf 8}, 343 (1959).

\bibitem{ssh} A. J. Heeger, S. A. Kivelson, J. R. Schrieffer, and W.-P. Su, Solitons in conducting polymers, Rev. Mod. Phys. {\bf  60}, 781 (1988).

\bibitem{ssh1} K. Hannewald, V. M. Stojanovi\'c, J. M. T. Schellekens, P. A. Bobbert, G. Kresse and J. Hafner,
Theory of polaron bandwidth narrowing in organic molecular crystals,
Phys. Rev. B {\bf 69}, 144302 (2004).

\bibitem{ssh2}
V. M. Stojanovic, and M. Vanevi\'c, Quantum-entanglement aspects of polaron systems, Phys. Rev. B {\bf 78}, 214301 (2008).

\bibitem{ssh3}
D. J. J. Marchand, G. De Filippis, V. Cataudella, M. Berciu, N. Nagaosa, N. V. Prokof'ev, A. S. Mishchenko, and P. C. E. Stamp, Sharp Transition for Single Polarons in the One-Dimensional Su-Schrieffer-Heeger Model,
Phys. Rev. Lett. {\bf 105}, 266605 (2010).

\bibitem{ssh4}F. Herrera, K. W. Madison, R. V. Krems, and M. Berciu, Investigating Polaron Transitions with Polar Molecules, Phys. Rev. Lett. {\bf 110}, 223002 (2013).


\bibitem{tian} V. M. Stojanovi\'c, M. Vanevi\'c, E. Demler, and L. Tian, Transmon-based simulator of nonlocal electron-phonon coupling: a platform for observing sharp small-polaron transitions, arXiv:1401.4783.


\bibitem{zagoskin} A. M. Zagoskin, S. Savel'ev, and F. Nori, Modeling an adiabatic quantum computer,
Phys. Rev. Lett. {\bf 98}, 120503 (2007).


\bibitem{geordie} M. W. Johnson, M. H. S. Amin,	S. Gildert,	T. Lanting,	F. Hamze, N. Dickson,	
R. Harris, A. J. Berkley, J. Johansson,	P. Bunyk, E. M. Chapple, C. Enderud, J. P. Hilton, K. Karimi,	
E. Ladizinsky, N. Ladizinsky, T. Oh, I. Perminov, C. Rich, M. C. Thom,	
E. Tolkacheva, C. J. S. Truncik, S. Uchaikin, J. Wang,	B. Wilson, and G. Rose,
Quantum annealing with manufactured spins, Nature
{\bf 473}, 194 (2011).

\bibitem{protein} A. Perdomo-Ortiz, N. Dickson, M. Drew-Brook, G. Rose, and  A. Aspuru-Guzik, Finding low-energy conformations of lattice protein models by quantum annealing, Sci. Rep. {\bf 2}, 571  (2012).

\bibitem{troyer} S. Boixo, T. F. R\o nnow, S. V. Isakov, Z. Wang, D. Wecker, D. A. Lidar, J. M. Martinis, and M. Troyer, Quantum annealing with more than one hundred qubits, arXiv:1304.4595.

\bibitem{troyerlidar}T. F. R\o nnow, Z. Wang, J. Job, S. Boixo, S. V. Isakov, D. Wecker, J. M. Martinis, D. A. Lidar, and M. Troyer, Defining and detecting quantum speedup, arXiv:1401.2910.

\bibitem{lidar} K. L. Pudenz, T. Albash, and D. A. Lidar, Error corrected quantum annealing with hundreds of qubits,
arXiv:1307.8190.

\bibitem{Tavis}  M. Tavis, and F. W. Cummings, Exact solution for an
n-molecule-radiation-field hamiltonian,  Phys. Rev. {\bf 170}, 379 (1968).

\bibitem{fink} J. M. Fink,
Dressed Collective Qubit States and
the Tavis-Cummings Model in Circuit QED, Phys. Rev. Lett. {\bf 103}, 083601 (2009).

\bibitem{vedral1}  G. S. Paraoanu, Realism and single-quanta nonlocality, Found. Phys. {\bf 41}, 734 (2011).

\bibitem{vedral2} L. Heaney, A. Cabello, M. F. Santos, and V. Vedral, Extreme nonlocality with one photon, New Journal of Physics {\bf 13}, 053054 (2011).

\bibitem{ustinov} P. Macha, G. Oelsner, J.-M. Reiner, M. Marthaler, S. Andr\'e, G. Sch\"on, U. Huebner, H.-G. Meyer, E. Il'ichev, A. V. Ustinov, Implementation of a Quantum Metamaterial, arxiv:1309.5268.


\end{thebibliography}
\end{document}